\documentclass{amsart}

\usepackage{graphicx}
\usepackage{orcidlink}

\begin{document}

\title{Merging Point Data for InSAR Deformation Processing}

\author{Matthew T. Calef\, \orcidlink{0000-0003-4701-7224}}
\email{mattcalef@descarteslabs.com}

\author{Kelly M. Olsen\, \orcidlink{0000-0002-2709-9237}}
\email{kelly.olsen@descarteslabs.com}

\author{Piyush S. Agram\, \orcidlink{0000-0003-0711-0264}}
\email{piyush@descarteslabs.com}

\begin{abstract}
  Given a collection of points $S \subset \mathbb{R}^N$, which is partitioned into $M$ overlapping subsets $\{S_i\}_{i=1}^M$, and approximate data $\{D_i\}_{i=1}^M$ associated with the subsets, one may seek a consistent merged dataset $D$ that is derived from $\{S_i\}_{i=1}^M$ and $\{D_i\}_{i=1}^M$. This note presents a method for constructing $D$ under the assumption that $D$ represents discrete samples of a suitably smooth function $f:\mathbb{R}^N \rightarrow \mathbb{R}$ evaluated at the points in $S$. The method has two steps. The first step uses a least-squares solve to approximate the constant offsets for each $D_i$. The second step uses a sequence of discrete Dirichlet problems to resolve any remaining differences. We include a two dimensional example of this method applied to deformation measurements derived from Interferometric Synthetic Aperture Radar (InSAR). 
\end{abstract}

\maketitle

\section{Introduction}

Given spatial datasets on overlapping point clouds, where the data are inexact, one may want to construct a dataset associated with the union of the point clouds. Interferometric synthetic aperture radar (InSAR)~\cite{Rosen:2000} provides a motivating example.

Synthetic Aperture Radar (SAR) is a remote-sensing technique in which an area is illuminated with a series of microwave pulses and an image is formed based on the microwaves that are scattered back to a sensor, which is often also the microwave emitter~\cite{Wiley:1985}. A SAR image is a two dimensional dataset consisting of complex numbers. Each complex number corresponds to a location within the area imaged. The amplitude of the complex number represents how strongly the location reflected the microwave pulses. The phase is determined by the scattering processes at the location and the distance traveled by the microwave pulses.

A central idea in InSAR is if two SAR images are collected over the same area and the scattering process does not change\footnote{This condition requires that images are collected from similar imaging geometries.}, the phase difference in the two images reflects the change in round trip distance between the two collection times. This allows one to recover surface deformation with an accuracy of a fraction of a wavelength, which is typically a few centimeters. To do so requires addressing two challenges. First, one needs to determine where the scattering process was nearly constant. These locations are called \emph{good points}. One such approach for selecting good points is described in~\cite{Ferretti:2001}. Second, the data at these good points is the change in round-trip distance between the two collection times \emph{modulo the wavelength}.

The problem of recovering the round-trip distance from the available data is referred to as \emph{unwrapping}. Unwrapping is ill-posed without additional assumptions about the smoothness of the deformation being measured. There are many approaches to unwrapping including~\cite{Goldstein:1988}, \cite{Chen:2001}, \cite{Pepe:2006}, and~\cite{Costantini:2011}. The computational cost of many of these unwrapping methods scales super-linearly with number of points, e.g. solving the linear programming problem in~\cite{Costantini:2011}, leading to an approach where data are processed in spatial partitions.

Unwrapping is similar to integrating an approximation of a gradient field to recover a potential, where the approximation of the gradient field may not be irrotational~\cite{Chartrand:2019}. As such unwrapping is a \emph{non local} problem and depends on all input data in the domain being unwrapped. In particular the unwrapped result at a point may differ depending on the surrounding input data used to generate that result.

The unwrapped data are estimates of the surface deformation between two points in time. One can recover per-epoch estimates of the cumulative deformation by solving a linear system, e.g. as described in~\cite{Berardino:2002}. Often one adds additional steps of using spatial and temporal trends to identify and remove any atmospheric effects whose phase contributions added artifacts to the estimated deformation.

The relevant properties of this problem are:
\begin{itemize}
  \item [1.] We seek to recover results $D \subset \mathbb{R}$ for a set of points $S \subset \mathbb{R}^2$, where $D$ is assumed to agree with a smooth function $f:\mathbb{R}^2 \rightarrow \mathbb{R}$ on $S$. In our example $S$ is the set of good points and $f$ is the deformation, which we assume to be spatially smooth. Each $D_i$ is our estimate of the cumulative deformation for the points in the spatial partition $S_i$.
  \item [2.] The partition of $S$, $\{S_i\}_{i=1}^M$, has the following property: if we form a graph with a vertex for each $S_i$ and edges between $i$ and $j$ for each $S_i \cap S_j \ne \emptyset$, the graph has one connected component. In our example $\{S_i\}_{i=1}^M$ reflects how we've partitioned the larger deformation estimation task into smaller tasks.
  \item [3.] For each partition $S_i$, the data $D_i$ that have artifacts associated with $S_i$. In our example each $D_i$ is the deformation estimate based on imperfectly unwrapping the data within $S_i$.
  \item [4.] The choice of the partition $\{S_i\}_{i=1}^M$ is arbitrary and should not be reflected in the merged data $D$. In our example, surface deformation clearly does not depend on how the computational problem is partitioned. If the final merged deformation estimate shows the partition boundaries, the merging process is deficient. 
\end{itemize}

\section{The Method}

The first step is to find approximate offsets for the data associated with each partition. We form a linear system that has as many columns as partitions.  For each pair of partition indices $i, j$ where $i < j$ and $S_i \cap S_j \ne \emptyset$, we compute the mean of $D_i - D_j$ on the overlap $S_i \cap S_j$. This mean becomes an entry on the right hand side of a linear system. The corresponding row has a $-1$ at column $i$ and a $1$ at columns $j$. This resulting matrix of differences has a null space consisting of constants per connected component. From property 2 above, there is only one connected component. A least-squares solution to this linear system provides approximate constant offset corrections which are applied to each $D_i$.

Due to property three, the corrected $D_i$ may still not agree on the overlapping regions. Assuming no $D_i$ is more accurate than another, one could average the $D_i$ on the overlapping regions. In practice this leads to jumps between regions that overlap to different degrees. Properties one and four implies these jumps indicate a poor solution.

Step two addresses these jumps. Each $S_i$ is made into a connected graph with a set of edges connecting nearby points, e.g. with edges from a Delaunay triangulation. Every point in $S$ is assigned an overlap degree, which is a count of the number of partitions to which that point belongs. Property two ensures every $S_i$ contains at least one point with overlap degree greater than one.

The method proceeds by computing the per-point average over the $D_i$ for the points with the highest overlap degree. It may be that not all $S_i$ contain a point of the highest overlap degree. For each $S_i$ that contains at least one point of highest overlap degree, one computes the per-point difference between the average and the values within the corresponding $D_i$ for the points with highest overlap degree.

This difference is the correction one would need to apply to the $D_i$ to ensure that they agrees with the average, however this correction is only available for the points that have the highest overlap degree.

The graph Laplacian associated with the graph on $S_i$ can be used to create a discrete Dirichlet problem where the boundary conditions are the correction values for the points of $S_i$ that have highest overlap degree. The solution to this problem is a correction to all of $D_i$ that has two properties. First, the solution agrees with the correction on the points of highest overlap degree. Second, the solution is harmonic with respect to the graph Laplacian on the rest of $S_i$, that is, the correction does not have local extrema and is, in some sense, smooth.

Applying the correction to each of the relevant $D_i$ ensures that the $D_i$ agree on the points with the highest overlap degree, and that the correction to achieve this smoothly extends from the points with highest overlap degree.

One then computes the correction to ensure the $D_i$ agree with the average values for points with the highest or second highest overlap degree. For relevant $D_i$, this correction is used as boundary conditions for another set of Dirichlet problems, where the boundary conditions are the points with highest or second highest overlap degree. The solutions to this second set of discrete Dirichlet problems is applied to the $D_i$.

Repeating this process to eventually include points with overlap degree two provides a solution to the problem described in the introduction.

\section{Implementation Details}

There are several opportunities for parallelism in this method. Given a worker per partition, where each worker can read from data associated with other workers, and can write to data associated with its own partition, this method can be implemented in the steps below. Note that the method does not proceed to a new step until all workers have completed the current step.

\begin{itemize}
    \item[1.] Worker $i$ computes a graph Laplacian using the links from a Delaunay triangulation of $S_i$. Worker $i$ then iterates over $j \ne i$ to compute overlap degrees and create maps between point sets. Worker $i$ computes the mean of $D_i - D_j$ on the overlap $S_i \cap S_j$, for all $j > i$.
    \item[2.] Worker $1$ gathers the values for the means of differences, forms and solves the least-squares problem to get the estimated constant offsets.
    \item[3.] Worker $i$ reads the constant offsets from worker $1$ and applies them.
    \item[4.] For every overlap degree, $P$, greater than one, in descending order, do the following:
    \begin{itemize}
        \item [] Worker $i$ reads from overlapping partitions and computes the mean over all overlapping partitions for points with the overlap degree $P$ or higher. Worker $i$ then computes the difference between this mean and $D_i$ for relevant points. Worker $i$ solves the Dirichlet problem that extends this difference smoothly from points with overlap degree $P$ or higher to all of $D_i$, and then applies this correction to $D_i$.
    \end{itemize}
\end{itemize}

Note that the synchronization between iterations of 4 can be relaxed, and this step can proceed for a worker at overlap degree $P$ so long as workers for overlapping partitions have completed the correction for overlap degree $P+1$. 

\section{Example}

We apply this merging method to InSAR deformation results over Mexico City. The SAR imagery were collected by Sentinel-1 along orbit track 78 between March 28, 2023 and April 3, 2024. We used the second method (Method 2) described in Section 4 of~\cite{Olsen:2023} to create estimates of deformation from complex SAR imagery on a partition of input data. The processing pipelines for coregistering the complex SAR imagery are described in~\cite{Agram:2022} and the compute environment is described in~\cite{Beneke:2017}.

\begin{figure}
\centering
\includegraphics[width=\linewidth]{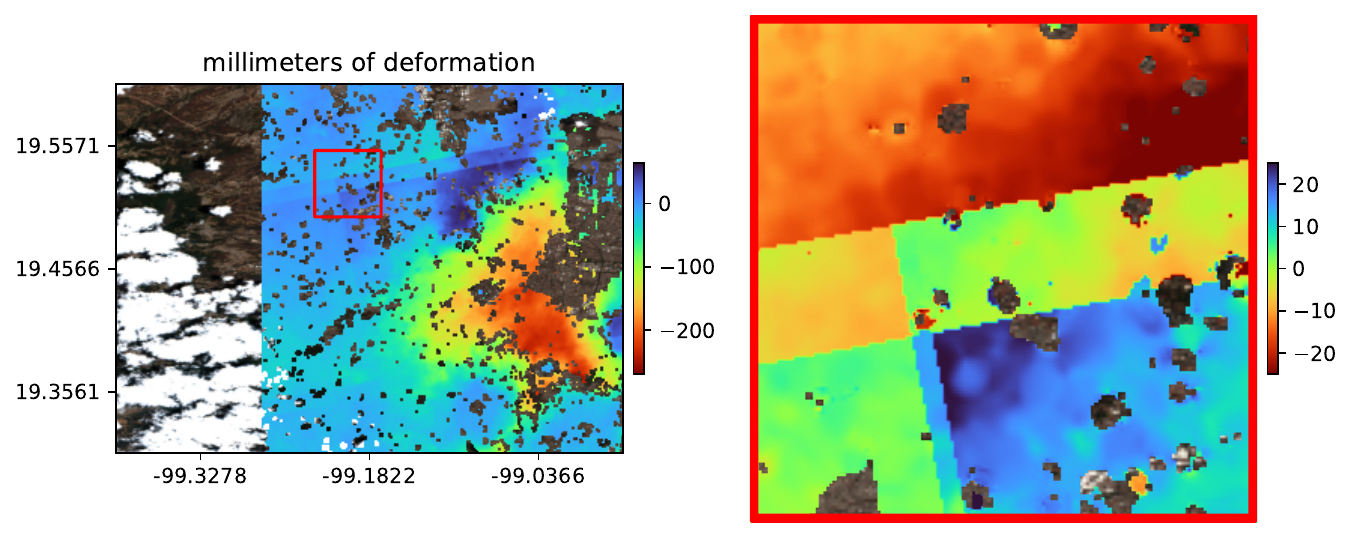}
\caption{The left image shows cumulative deformation averaged over the spatial partitions. The area within the red rectangle is shown in detail on the right. Note that the color scale on the right is compressed to show clearly the discrepancies in the deformation in the overlapping regions.}
\label{fig:raw}
\end{figure}

\begin{figure}
\centering
\includegraphics[width=\linewidth]{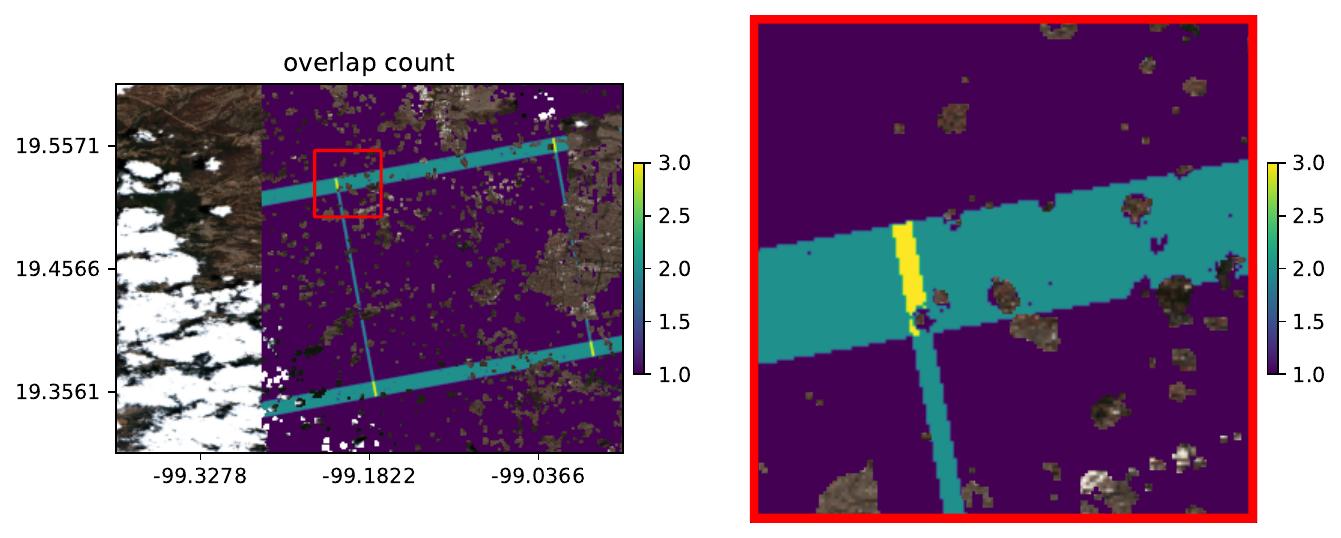}
\caption{The left image shows the count of overlapping partitions per point selected by each partition. The area within the red rectangle is shown in detail on the right. Yellow indicates three results overlap, teal indicates two, and dark purple indicates no overlap.}
\label{fig:overlaps}
\end{figure}

Figure~\ref{fig:raw} shows the cumulative deformation averaged from five overlapping deformation results. Figure~\ref{fig:overlaps} shows the degree of overlap. The deformation estimation provides results up to an unknown additive constant for each partition. The non-local nature of unwrapping methods often leads to unwrapped results that differ by more than a constant on overlapping regions.

\begin{figure}
\centering
\includegraphics[width=\linewidth]{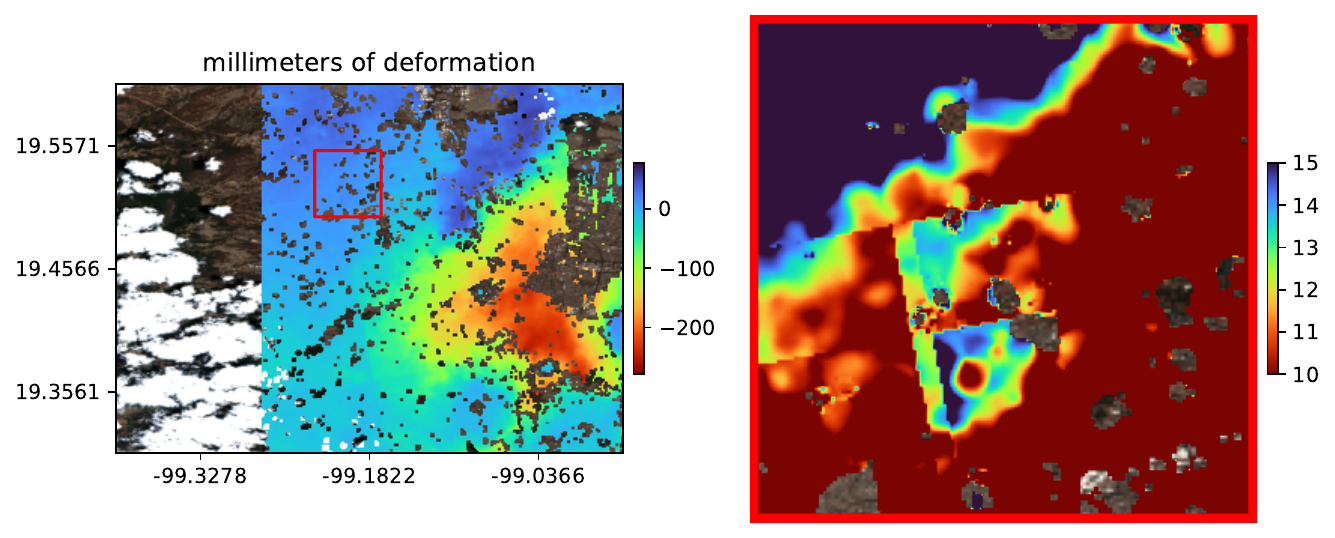}
\caption{The left image shows average deformation results after computing a least-squares approximation for the constant offsets. The detail on the right shows the remaining deformation differences after correcting by a constant. Note that the color scale has been further compressed compared to the color scale on the right of Figure~\ref{fig:raw}} 
\label{fig:merged0}
\end{figure}

This remaining difference after estimating and removing these per-partition constants is clear in Figure~\ref{fig:merged0}. Note that applying the least-squares approximation of the constant offsets leaves artifacts of roughly 5mm in size. Addressing these remaining artifacts are accomplished by solving two sets of Dirichlet problems as described above.

For each partition, we compute the difference between the average over all partitions and the value for the specific partition for points with overlap degree three. We solve the discrete graph-Laplace equation with this difference for points with overlap degree three (the highest overlap degree) as a boundary condition. The solution to this problem agrees with the difference on the boundary condition, and has no local maxima or minima extending away from the the degree three overlap points.

\begin{figure}
\centering
\includegraphics[width=\linewidth]{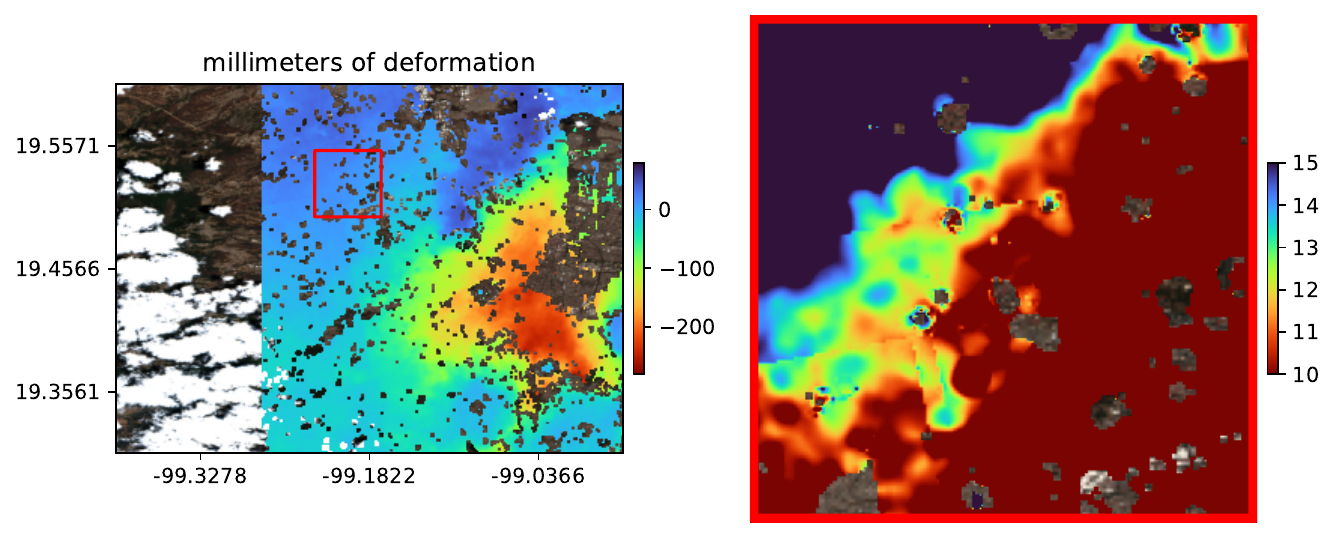}
\caption{The left image shows average deformation results after applying the correction from the first set of Dirichlet problems. The red square is shown in detail on the right. 
} 
\label{fig:merged1}
\end{figure}

Applying the solution to this first Dirichlet problem leads to deformation estimates that agree on the regions with overlap degree three, and show no artifacts along the edges of the overlap three regions. One can see in the right panel of Figure~\ref{fig:merged1} that there are no remaining artifacts along the edges of the yellow region shown in the right hand panel of Figure~\ref{fig:overlaps}.

\begin{figure}
\centering
\includegraphics[width=\linewidth]{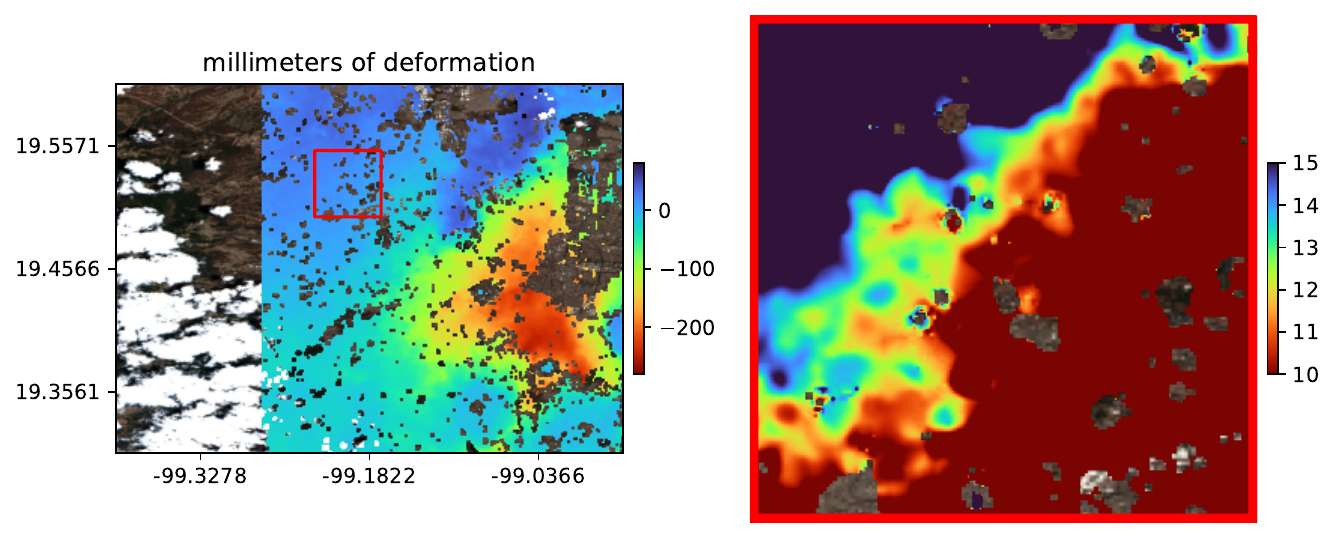}
\caption{The left image shows average deformation results after applying the correction from the second set of Dirichlet problems. The red square is shown in detail on the right.} 
\label{fig:merged2}
\end{figure}

We repeat this process to extend the difference between the average over all the partitions and the per-partition values for points with overlap degree two or higher. Applying this correction ensures that data for each partition agree with the average on points with overlap degree two, and extend smoothly away from points with overlap degree two. The result is shown in Figure~\ref{fig:merged2}. This completes the merging method.

\section{Discussion}

\subsection{Computational Considerations}

The graph Laplacian associated with each $S_i$ can be reused for the sequence of Dirichlet problems. In fact the graph Laplacians can be reused for multiple merging processes, so long as the underlying $S_i$ don't change.

The Delaunay triangulations for each $D_i$ may not agree in the regions where the $D_i$ overlap. We have not found this to lead to artifacts.

We used the python SciPy package~\cite{scipy:2020} for sparse matrix data structures and for the Delaunay triangulation. We used conjugate gradient within SciPy to solve the Dirichlet problems.

The five input partitions had the following numbers of points 139621, 150160, 106318, 235506, 36819. The final merged dataset had 613194 points. The merging was performed on a single core of an AMD Epyc 7571. The compute times are shown in Table~\ref{tab:times}

\begin{table}
  \caption{These are times to perform the steps of the merging method}
  \begin{tabular}{ll}
    Operation & Time \\
    \hline \\
    Computing overlaps & 104ms \\
    Computing least squares offsets & 6.28ms \\
    Forming the graph Laplacians as sparse matrices & 62s \\
    Solving the sequence of Dirichlet problems & 43s
  \end{tabular}
  \label{tab:times}
\end{table}

It is our experience that the limiting factor is memory, not compute time. A straightforward implementation would bring all the data to be merged into a common memory space, and should be avoided.

\subsection{Algorithmic Considerations}

The European Ground Motion Service (EGMS) has a goal of large scale deformation monitoring via InSAR, and it identifies the need for spatial merging. The EGMS report~\cite{Larsen:2020} recognizes several relevant works. The wide area persistent scatterer (PS) interferometry approach~\cite{Adam:2011} proposes to estimate error propagation for a comparatively low density set of PS points. Manunta et al~\cite{Manunta:2019} propose a parallel SBAS approach, while Costantini et al~\cite{Costantini:2011} uses linear or quadratic programming to ensure consistency in unwrapped results. In these three methods, merging is part of the InSAR processing chain. In contrast the method we propose is quite general and is independent of the underlying InSAR deformation retrieval method.

This method has an issue worth noting. Many unwrapping methods produce an unwrapped phase that differs from input phase by integer multiples of $2\pi$, e.g. those described in~\cite{Costantini:2011} and~\cite{Pepe:2006}. One might want the merging process to respect this integer difference property. That is to say, the merged unwrapped phase should continue to differ from the wrapped phase by integer multiples of $2\pi$. The method described in this note does not ensure this property, while the tiling approach in, e.g., the SNAPHU unwrapper~\cite{Chen:2002} does maintain this property. 

In the example we presented, we applied the proposed merging method to deformation estimates. These deformation estimates are based on unwrapped interferograms, but also add corrections that contribute fractions of a $2\pi$ cycle. These corrections include removing estimated atmospheric effects and estimated phase contributions from errors in the digital elevation model. Since the input to this merge process already differs from the wrapped phase by fractions of a $2\pi$ cycle, there is no need not preserve the integer difference with the input. However, we recognize the desire for this property when applying merging directly to unwrapped interferometric phases.

\subsection{Next steps}

We aim to adapt this method to ensure that the phase correction associated with merging can be restricted to integer multiples of $2\pi$, thereby addressing concerns above.

\bibliographystyle{unsrt}
\bibliography{refs}

\begin{thebibliography}{10}

\bibitem{Rosen:2000}
P.~A. {Rosen}, S.~{Hensley}, I.~R. {Joughin}, F.~K. {Li}, S.~N. {Madsen}, E.~{Rodriguez}, and R.~M. {Goldstein}.
\newblock Synthetic aperture radar interferometry.
\newblock {\em Proceedings of the IEEE}, 88(3):333--382, 2000.

\bibitem{Wiley:1985}
Carl~A Wiley.
\newblock Synthetic aperture radars.
\newblock {\em IEEE Transactions on Aerospace and Electronic Systems}, (3):440--443, 1985.

\bibitem{Ferretti:2001}
A.~{Ferretti}, C.~{Prati}, and F.~{Rocca}.
\newblock Permanent scatterers in sar interferometry.
\newblock {\em IEEE Transactions on Geoscience and Remote Sensing}, 39(1):8--20, 2001.

\bibitem{Goldstein:1988}
Richard~M Goldstein, Howard~A Zebker, and Charles~L Werner.
\newblock Satellite radar interferometry: Two-dimensional phase unwrapping.
\newblock {\em Radio science}, 23(4):713--720, 1988.

\bibitem{Chen:2001}
Curtis~W. Chen and Howard~A. Zebker.
\newblock Two-dimensional phase unwrapping with use of statistical models for cost functions in nonlinear optimization.
\newblock {\em J. Opt. Soc. Am. A}, 18(2):338--351, Feb 2001.

\bibitem{Pepe:2006}
Antonio Pepe and Riccardo Lanari.
\newblock On the extension of the minimum cost flow algorithm for phase unwrapping of multitemporal differential sar interferograms.
\newblock {\em IEEE Transactions on Geoscience and remote sensing}, 44(9):2374--2383, 2006.

\bibitem{Costantini:2011}
Mario Costantini, Fabio Malvarosa, and Federico Minati.
\newblock A general formulation for redundant integration of finite differences and phase unwrapping on a sparse multidimensional domain.
\newblock {\em IEEE Transactions on Geoscience and Remote Sensing}, 50(3):758--768, 2011.

\bibitem{Chartrand:2019}
R.~{Chartrand}, M.~T. {Calef}, and M.~S. {Warren}.
\newblock Exploiting sparsity for phase unwrapping.
\newblock In {\em IGARSS 2019 - 2019 IEEE International Geoscience and Remote Sensing Symposium}, pages 258--261, 2019.

\bibitem{Berardino:2002}
P.~{Berardino}, G.~{Fornaro}, R.~{Lanari}, and E.~{Sansosti}.
\newblock A new algorithm for surface deformation monitoring based on small baseline differential sar interferograms.
\newblock {\em IEEE Transactions on Geoscience and Remote Sensing}, 40(11):2375--2383, 2002.

\bibitem{Olsen:2023}
Kelly~M Olsen, Matthew~T Calef, and Piyush~S Agram.
\newblock Contextual uncertainty assessments for insar-based deformation retrieval using an ensemble approach.
\newblock {\em Remote Sensing of Environment}, 287:113456, 2023.

\bibitem{Agram:2022}
Piyush~S Agram, Michael~S Warren, Matthew~T Calef, and Scott~A Arko.
\newblock An efficient global scale sentinel-1 radar backscatter and interferometric processing system.
\newblock {\em Remote Sensing}, 14(15):3524, 2022.

\bibitem{Beneke:2017}
Carly~Marie Beneke, Samuel Skillman, Michael~S Warren, Tim Kelton, Steven~Patrick Brumby, Rick Chartrand, and Mark Mathis.
\newblock A platform for scalable satellite and geospatial data analysis.
\newblock In {\em AGU Fall Meeting Abstracts}, volume 2017, pages IN32C--04, 2017.

\bibitem{scipy:2020}
Pauli Virtanen, Ralf Gommers, Travis~E. Oliphant, Matt Haberland, Tyler Reddy, David Cournapeau, Evgeni Burovski, Pearu Peterson, Warren Weckesser, Jonathan Bright, St{\'e}fan~J. {van der Walt}, Matthew Brett, Joshua Wilson, K.~Jarrod Millman, Nikolay Mayorov, Andrew R.~J. Nelson, Eric Jones, Robert Kern, Eric Larson, C~J Carey, {\.I}lhan Polat, Yu~Feng, Eric~W. Moore, Jake {VanderPlas}, Denis Laxalde, Josef Perktold, Robert Cimrman, Ian Henriksen, E.~A. Quintero, Charles~R. Harris, Anne~M. Archibald, Ant{\^o}nio~H. Ribeiro, Fabian Pedregosa, Paul {van Mulbregt}, and {SciPy 1.0 Contributors}.
\newblock {{SciPy} 1.0: Fundamental Algorithms for Scientific Computing in Python}.
\newblock {\em Nature Methods}, 17:261--272, 2020.

\bibitem{Larsen:2020}
Y~Larsen, P~Marinkovic, JF~Dehls, M~Bredal, C~Bishop, G~J{\o}kulsson, LP~Gj{\o}vik, R~Frauenfelder, SE~Salazar, M~V{\"o}ge, et~al.
\newblock European ground motion service: Service implementation plan and product specification document.
\newblock {\em Copernicus Land Monitoring Service}, 2020.

\bibitem{Adam:2011}
Nico Adam, Fernando~Rodriguez Gonzalez, Alessandro Parizzi, and Werner Liebhart.
\newblock Wide area persistent scatterer interferometry.
\newblock In {\em 2011 IEEE International Geoscience and Remote Sensing Symposium}, pages 1481--1484. IEEE, 2011.

\bibitem{Manunta:2019}
Michele Manunta, Claudio De~Luca, Ivana Zinno, Francesco Casu, Mariarosaria Manzo, Manuela Bonano, Adele Fusco, Antonio Pepe, Giovanni Onorato, Paolo Berardino, et~al.
\newblock The parallel sbas approach for sentinel-1 interferometric wide swath deformation time-series generation: algorithm description and products quality assessment.
\newblock {\em IEEE Trans. Geosci. Remote. Sens.}, 57(9):6259--6281, 2019.

\bibitem{Chen:2002}
Curtis~W Chen and Howard~A Zebker.
\newblock Phase unwrapping for large sar interferograms: Statistical segmentation and generalized network models.
\newblock {\em IEEE Transactions on Geoscience and Remote Sensing}, 40(8):1709--1719, 2002.

\end{thebibliography}

\end{document}